# Integrating Spatial and Temporal Effects in Seat-Belt Compliance Assessment with Telematics Data

**Ashutosh Dumka (Corresponding Author)**
Graduate Student
Department of Civil Construction and Environmental Engineering,
Iowa State University of Science and Technology
Ames, IA 50011-1066, USA
E-mail: adumka@iastate.edu

**Raghupathi Kandiboina,** PhD
Department of Civil Construction and Environmental Engineering,
Iowa State University of Science and Technology
Ames, IA 50011-1066, USA
E-mail: raghukan@iastate.edu

**Skylar Knickerbocker**
Research Scientist
Institute for Transportation,
Iowa State University of Science and Technology,
Ames, IA 50011-1066, USA
E-mail: sknick@iastate.edu

**Neal Hawkins**
Associate Director
Institute for Transportation,
Iowa State University of Science and Technology
Ames, IA 50011-1066, USA
E-mail: hawkins@iastate.edu

**Jonathan Wood**
Assistant Professor
Department of Civil Construction and Environmental Engineering,
Iowa State University of Science and Technology
Ames, IA 50011-1066, USA
E-mail: jwood2@iastate.edu

**Anuj Sharma**
Professor
Department of Civil Construction and Environmental Engineering,
Iowa State University of Science and Technology
Ames, IA 50011-1066, USA
E-mail: anujs@iastate.edu

**ABSTRACT**

Seat belt use remains one of the most effective measures for reducing vehicle occupant fatalities and injuries. Yet, seat-belt compliance across different locales demands far more granular data than traditional, roadside surveys can provide. These surveys are spatially sparse, temporally intermittent, and costly to administer, often providing coarse-grained snapshots insufficient for capturing dynamic behavioral patterns or localized disparities. Telematics data emerges as a transformative alternative, offering continuous, high-resolution "driver event" records, such as seat belt latch status, across vast geographic areas. This granular and scalable data enables the application of advanced spatiotemporal models that more accurately reflect the complex interactions driving seatbelt use. This study utilizes telematics data to generate county-level seat belt compliance metrics for Iowa in 2022, employing a suite of beta-regression models that incorporate spatial and temporal random effects. The study's findings demonstrate that models including both spatial and temporal components outperform those with spatial or temporal effects alone, underscoring the importance of jointly accounting for geographic clustering and temporal dynamics. Among explanatory variables, vehicle miles traveled (VMT) and per capita income emerge as significant predictors of compliance rates. The significant spatial and temporal effects highlight that telematics-based granular data substantially enhances model fit and inference quality. The results demonstrate that integrating granular telematics data within sophisticated spatiotemporal frameworks significantly improves inference, providing policymakers with precise insights for targeted interventions and advancing traffic safety research.

**INTRODUCTION AND LITERATURE REVIEW**

Seat belt use remains one of the most effective safety measures in vehicles, reducing fatality risk by roughly 45% and serious injury by about 50% in frontseat occupants [1]. Decades of legislation and public campaigns have driven national compliance to approximately 91% in 2022 [2], yet this aggregate figure masks persistent demographic and geographic disparities. While these surveys provide essential snapshots of belt-wearing rates, they cannot capture long-term trends or the complex interplay of demographic, policy, and environmental factors that influence compliance over time and across different contexts. To evaluate the sustained impact of safety interventions and identify emerging compliance disparities, we require analytical frameworks, specifically statistical models that explicitly incorporate both spatial and temporal dependencies to monitor belt use continuously, disentangle latent drivers of change, and inform more targeted, equity-focused campaigns.

Despite their longstanding role in traffic-safety research, manual roadside observational surveys struggle to capture compliance disparities across diverse populations and locations. Meta-analyses and extensive surveys consistently report lower belt use among men, younger adults (18-34 years old), rear-seat occupants, individuals with lower incomes, and residents of secondary enforcement jurisdictions [3]. Rural areas and specific demographics have lagged in seat belt use; for example, a 2014 study found that self-reported seat belt use was around 75% in rural U.S. counties, compared to ~89% in urban counties [4]. Traditional methods for measuring seat belt use, primarily manual roadside observational surveys, have historically served as the foundation for estimating compliance [5, 6]. However, these surveys face inherent constraints, including limited spatial and temporal coverage, potential sampling biases, and substantial logistical and cost burdens. Observations are often conducted at fixed sites and specific times, which restricts their representativeness and the ability to capture dynamic changes or localized behaviors in seat belt usage. Furthermore, these surveys typically provide aggregate snapshots without the granularity required to disentangle the complex interactions between demographic characteristics, enforcement policies, and situational variables that drive compliance. In Iowa, for example, the annual statewide seat belt survey samples 84 fixed sites across 15 counties to estimate compliance [7]. Efforts to improve upon the spatial and temporal constraints of manual roadside surveys have led researchers to explore unmanned aerial systems as an alternative means of observing belt-wearing behavior. Tethered drones can hover over roadways to record high-resolution video for direct compliance assessment, reducing the need for human observers at each location and allowing rapid deployment across varied settings [8]. However, their operational range and sampling coverage are still restricted by endurance, line-of-sight, and logistical constraints, resulting in spatial sparsity and inherent sampling limitations. These operational constraints motivate the search for continuous, scalable behavioral data, a need addressed by telematics data streams.

Measuring the real-world impact of these interventions is critical, as improved compliance not only translates to fewer fatalities but also reflects the effectiveness of policy decisions and the allocation of resources. To comprehensively assess how sociodemographic and environmental factors influence seat belt use, emerging research advocates for leveraging richer datasets that offer finer geographic and temporal resolution combined with advanced analytical frameworks. Novel data sources, such as telematics, provide continuous event-level information across extensive spatial scales, enabling measurement beyond sparse sampled locations. In principle, telematics data allow for a much richer spatio-temporal picture: they can generate county-level or even road-specific seatbelt compliance rates on a daily or hourly basis, far beyond the temporal granularity of manual surveys. Early work has shown that probe vehicle datasets can feasibly capture traffic behaviors at scale, albeit with a modest market penetration of only a few percent of vehicles. For instance, a recent analysis of telematics data in Iowa found an average penetration of roughly 6.3%, yet still recorded massive volumes of trips and demonstrated reliable speed measurements [8]. Furthermore, studies have shown the effectiveness of using telematics data in real-world applications for detecting lane closures and real-time traffic crash detection [9, 10]. This indicates that even a small fraction of instrumented vehicles can provide valuable information, covering a broad geographic footprint at low cost. However, raw counts of telematics latch events alone cannot reveal the unobserved regional and seasonal factors driving seat-belt compliance.

In summary, overcoming the limitations of manual surveys and adequately addressing the multidimensional heterogeneity of seat-belt compliance requires the integration of high-resolution data sources with advanced spatiotemporal modeling techniques. These integrative frameworks are essential for uncovering the subtle, context-specific drivers of compliance disparities and designing targeted, equity-focused interventions. Research has examined spatial correlations in seatbelt non-use, finding that factors beyond internal structural factors influence this behavior. Additionally, the relationship between demographic variables and seatbelt non-use has been explored across different regions [11]. Another study conducted by [12] explores the variation in drivers' seatbelt use in relation to indicators of community-level factors. Although a handful of studies [13, 14] have conducted either spatial or temporal analyses of seat-belt use and related laws, comprehensive examinations that jointly model both spatial and temporal dependencies and systematically compare multiple model specifications remain virtually nonexistent. In particular, no study to date has explored how different combinations of spatial effects, temporal structures, and their interactions influence estimates of compliance across adjacent jurisdictions; hence, advanced statistical models are needed to study seatbelt use.

To disentangle the seatbelt usage and avoid biased inference, this study employs advanced spatiotemporal models that explicitly incorporate both spatial clustering and temporal dynamics. The heterogeneity in seat belt use, shaped by spatial dependencies where adjacent areas exhibit correlated behaviors due to shared socioeconomic or cultural attributes and temporal trends driven by policy changes or awareness campaigns, necessitates models capable of capturing spatio-temporal correlations. Complex statistical approaches, particularly hierarchical Bayesian models that incorporate spatial and temporal random effects, have demonstrated a superior capacity to parse these dependencies and improve estimation accuracy over traditional regression analyses conducted in isolation.

Spatial dependence in traffic safety has been extensively documented in crash frequency research, where geographically adjacent areas often share unobserved characteristics, such as local enforcement practices, land-use patterns, and infrastructure design, that drive the clustering of crash risk. [15, 16, 17]. At broad scales (e.g., counties or regions), Bayesian hierarchical models employing conditional autoregressive (CAR) and Besag-York-Mollié (BYM) priors have been shown to markedly enhance model fit and risk estimation by explicitly accounting for these latent spatial correlations [18, 19, 20]. Diagnostics such as Moran's I and local indicators of spatial association (LISA) further corroborate the presence of significant crash clusters, underscoring that neglecting spatial structure can lead to biased effect estimates and misidentification of high-risk zones. At finer scales, intersections and roadway segments studies using local Moran's I and Bayesian "corridor" models similarly demonstrate that crash occurrences are linked through roadway geometry, traffic flow characteristics, and corridor [21, 22, 23]. These analyses pinpoint persistent local hotspots and reveal how micro- and macro-level spatial processes interact across the transportation network. By analogy, seat-belt compliance can exhibit spatial patterns driven by shared socioeconomic, cultural, and enforcement contexts. Although few studies have formally modeled spatial or temporal dependencies in belt-wearing behavior, the robust findings from crash modeling provide a clear blueprint for future research.

Temporal correlation has long been central to crash modeling, with studies repeatedly demonstrating that crash frequencies and severities are not independent over time, but instead display pronounced cyclical and period-specific trends at yearly, monthly, weekly, daily, and even hourly scales [22, 24]. Influences such as weather, economic activity, law enforcement, and travel demand often underlie these temporal dynamics, producing distinct patterns, including higher crash rates during weekends, holidays, or adverse weather seasons. These patterns must be rigorously accounted for using methods like autoregressive time series models and spatio-temporal Bayesian frameworks. [25, 26, 27]. A parallel exists in seat belt compliance research, where usage rates exhibit temporal variability influenced by similar factors policy changes, public campaigns, and seasonal or situational behaviors. Several studies using cross-sectional time-series and event-level telematics have shown that belt use fluctuates by hour, day, and season, with compliance often declining during nighttime, weekends, or adverse conditions. Capturing this temporal autocorrelation is crucial for accurate estimation and targeting of interventions, as failing to account for it risks misattributing variability to other factors and understating both the challenge and the opportunity for improving road safety.

Several established statistical approaches have been widely discussed in the literature for modeling seatbelt compliance rates, which typically range from zero to one. The fractional logit model, introduced by Papke and Wooldridge, is particularly suitable for non-integer proportion data, as it employs a logistic link function and quasi-maximum likelihood estimation to model conditional means directly. It has been effectively applied in transportation contexts, such as mode share analysis, in the work of [28, 29]. Tobit models, commonly used for censored cases where values often cluster at the upper or lower limits, have been employed in traffic safety studies to address data truncation. However, they may introduce bias near the boundaries when observations are densely packed in that area [30]. Beta regression is exceptionally flexible and robust for strictly fractional variables (i.e., continuous outcomes within the interval (0, 1)), simultaneously modeling both the mean and the dispersion (precision) parameter [31].

In summary, traditional seat belt compliance studies face significant limitations. Manual roadside surveys are temporally and spatially sparse, often limited to a few sampled counties and fixed observation times, restricting the ability to capture fine-grained, dynamic compliance variations. Although emerging methods, such as UAV-based surveys, improve spatial coverage, they remain hindered by logistical constraints and are far from continuous. Moreover, most seat belt studies analyze spatial or temporal factors in isolation, lacking comprehensive models that integrate both dimensions simultaneously. This gap contrasts with the more advanced spatial-temporal modeling that has been extensively developed in traffic crash research, where frameworks such as Bayesian hierarchical models with CAR/BYM spatial priors and autoregressive temporal components have robustly characterized clustering and trends.

Building on the methodological foundation of spatial-temporal crash modeling, this study advances seat-belt compliance research through the following clear, staged contributions:

- *Utilization of High Resolution Telemaics Data*: Unlike traditional roadside surveys, this study leverages continuous, large scale telematics event data to capture granular spatial and temporal patterns of seat belt use
- *Spatio-Temporal Modeling Framework*: The study uses the application of hierarchical beta-regression models incorporating both spatial (CAR/BYM priors) and temporal (AR(1)/RW(1) processes) random effects.
- *Policy Relevant Insights on Compliance Drivers*: By integrating detailed telematics data with demographic and environmental covariates, the study identifies significant predictors. These findings provide policymakers with actionable insights to design targeted interventions and assess the effectiveness of enforcement and awareness campaigns.

The paper is structured as follows. Section 1 describes data sources and the derivation of county-level seat-belt compliance rates. Section 2 assesses spatial dependence using Moran's I and neighborhood diagnostics. Section 3 outlines the modeling framework, including spatial, temporal, and spatiotemporal beta regressions. Section 4 presents results and analysis of the best-fit model. Section 5 concludes with key findings and policy implications.

## DATA DESCRIPTION

The telematics data utilized in the study contains both "Vehicle Movement Data" and "Driver Events Data" categories [32]. To compute county-level seat-belt compliance rates, the telematics "Driver Events" dataset, which contains event tags for seat-belt latch, engine on/off, headlight on/off, etc., was merged with the "Vehicle Movement" dataset using the unique journey identifier. This linkage ensured that each movement record carried its associated latch events. A journey is then flagged as "latched" if it recorded at least one seat-belt-latch event at the beginning of the journey following engine-on, thereby focusing on latches occurring at or immediately after ignition and minimizing misclassification from pre-ignition buckling or mid-trip latchings. For each county, the number of unique journeys with an early latch event was counted and then divided by the total number of observed journeys. The resulting ratio, ranging from 0 to 1, represents the seat-belt compliance rate for that county, providing a consistent and conservative measure of actual belt use across Iowa.

However, evaluating whether this data accurately represents traffic patterns is crucial before implementing it. A study by [33] assessed the suitability of data for traffic applications, such as incident detection, signal optimization, and work zone management. During the evaluation, they focused on the penetration rate, which refers to the proportion of vehicles equipped with telematics compared to those equipped with fixed sensors. The findings showed a penetration rate of 6.3%, indicating that telematics captured 6.3% of vehicles on the road. Other studies on similar probe data during 2020 and 2021 resulted in a penetration rate between 4 to 5 percent in Ohio, Indiana, and Pennsylvania [33, 34]. While further research may be necessary to generalize these results across road types, this study highlights the promise of data as a valuable asset in modern transportation systems. The vehicle miles traveled (VMT) data for each county in each year, specifically for 2022, were downloaded from the Iowa Department of Transportation's website. In addition, some county level variables, such as Social, economic, and demographic characteristics of the counties in Iowa, were obtained from the US Census website [35]. A summary of the variables is shown in Table 1.

**Table 1 Descriptive Statistics of the Variables used in the study**

| Variables | Mean | Std. Dev | Median | Min | Max |
|---|---|---|---|---|---|
| Seatbelt Compliance (%) | 0.63 | 0.07 | 0.64 | 0.25 | 0.81 |
| Per capita income ($10,000) | 3.55 | 0.36 | 3.5 | 2.63 | 5.13 |
| Mean family income ($10,000) | 10.24 | 1.24 | 10.02 | 8 | 15.74 |
| Mean household income ($10,000) | 8.51 | 0.96 | 8.26 | 6.67 | 13.02 |
| BPL families (%) | 6.98 | 2.64 | 6.5 | 2.5 | 15.2 |
| Total household (per 1000) | 13.03 | 23.99 | 6.21 | 1.52 | 199.01 |
| Average family size | 2.96 | 0.14 | 2.96 | 2.73 | 3.47 |
| School Enrollment (per 1000) | 7.97 | 15.36 | 3.18 | 0.76 | 119.51 |
| Young population (%) | 0.08 | 0.02 | 0.08 | 0.05 | 0.21 |
| Middle-aged population (%) | 0.61 | 0.03 | 0.6 | 0.55 | 0.73 |
| Old population (%) | 0.31 | 0.04 | 0.32 | 0.18 | 0.39 |
| Male Population (%) | 0.51 | 0.01 | 0.5 | 0.48 | 0.53 |
| Female Population (%) | 0.49 | 0.01 | 0.5 | 0.47 | 0.52 |
| VMT (100,000 miles) | 3.33 | 4.96 | 1.94 | 0.5 | 41.41 |
| Unemployment rate (%) | 3.36 | 1.2 | 3.3 | 0.7 | 6.6 |

Note: VMT, vehicle miles traveled; BPL, below the poverty line

The Pearson correlation of the variables revealed a strong correlation between mean family income, per capita income, and mean household income, so per capita income was used in the further analysis. Additionally, no other variables showed a high correlation with each other. Thus, no other variables were removed for further analysis.

Unobserved heterogeneity caused by spatial and temporal correlation in data can often be identified by visualizing the data and further corroborated by statistical methods, such as Moran's I, which can help identify any spatial correlation. Figure 1 shows the plot of average seatbelt compliance for the counties in Iowa for the year 2022.

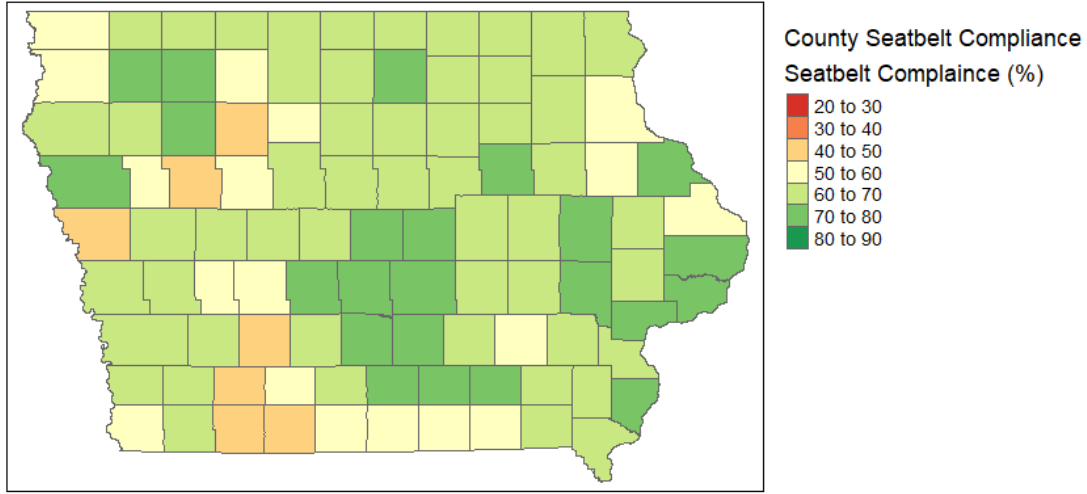

**Figure 1: County-level monthly average of seatbelt compliance percentage for Iowa (2022)**

The Pearson correlation of the variables revealed a strong correlation between average family income, per capita income, and mean household income; therefore, only per capita income was used in further analysis. Additionally, no other variables showed a high correlation with each other. Thus, no other variables were removed for further study.

**Strengths and Caveats of Telematics-Derived Compliance Metrics**
The use of telematics data presents an emerging opportunity for large-scale behavioral monitoring in transportation safety studies, including the analysis of seatbelt usage. Although the current penetration rate of telematics-equipped vehicles is approximately 6-7%, this figure represents the proportion of vehicles actively transmitting data, rather than the coverage or representativeness in terms of spatial reach. In practice, the presence and dispersion of telematics data spans nearly all road networks and counties, allowing for a much broader observational footprint than traditional manual surveys. Unlike the Iowa Seat Belt Use Survey, which samples 84 predefined road segments across 15 counties to generate weighted statewide compliance estimates, telematics data captures continuous, location-tagged behavioral data from a wide geographic base. This enables the generation of county-level, road-type-specific seatbelt compliance trends, typically on a daily or hourly basis. Each telematic strip provides journey-level context, including information on whether the seatbelt was engaged during travel, regardless of road classification. This significantly enhances the analysis resolution, especially for identifying spatial disparities and temporal trends in seatbelt use that may not be captured in limited sample manual surveys.

However, despite this scalability, telematics-based seatbelt data is not without limitations. Two key issues constrain its direct comparability with ground truth observational data: First, seatbelt usage is typically recorded only after engine ignition, which can result in underreporting if occupants fasten their seatbelts before starting the vehicle. Second, while spatially broad, a low penetration rate may introduce sample bias, particularly in rural areas or vehicle fleets with lower adoption of telematic technologies. A validation exercise was conducted at two selected locations using manual observation and UAV-based (drone) video analysis to assess the deviation. For these two locations in Story County, a video sampling was conducted for 4 hours using the UAV-based drone approach. The UAV-based computer vision method estimated seatbelt compliance for the story at 81%, which was higher than the telematics-derived estimate (~76%). This evidence confirms that while telematics data tends to underreport compliance slightly, the deviation remains within an acceptable range for trend monitoring. Taken together, telematics-based data should not be considered a replacement for statistically weighted manual surveys, particularly when precise compliance rates are needed for policy reporting. However, its advantages in geographic reach, temporal granularity, and passive scalability make it a powerful supplementary tool for evaluating broad spatial trends, uncovering areas of concern, and prioritizing locations for targeted enforcement or awareness campaigns. In summary, telematics data offers significant analytical potential in traffic safety. When used transparently and with acknowledgment of its limitations, it can play a crucial role in complementing traditional survey methods, bridging the gap between small sample accuracy and population-level visibility.

To observe the unobserved heterogeneity caused by the spatial and temporal correlation of data, we often visualize the data and then use statistical methods to corroborate our findings. The monthly average seatbelt usage percentage for each county in 2022 is shown in Figure 1. As expected, visual inspection reveals that seatbelt usage exhibits some clustering in the counties. The following statistical analysis was performed to investigate the presence of spatial correlation. Before computing spatial autocorrelation statistics, a spatial neighborhood structure was defined using the Queen contiguity criterion, where counties sharing either a border or a vertex are considered neighbors. Figure 2 displays the spatial connectivity map for Iowa counties, where red lines indicate the connections based on this contiguity. This structure was used to create the spatial weight matrix necessary for Moran's I test. Moran's statistic is commonly used to test the spatial correlation. The global Moran's I is defined as:

$$I = \frac{n \Sigma_i \Sigma_i \omega_{ij}(y_i - \bar{y})(y_j - \bar{y})}{\Sigma_{i \neq j} \omega_{ij} \Sigma_i (y_i - \bar{y})^2}$$

Where n is the total number of observations, $y_i$ and $y_j$ are the values of observation I and observation j, $\bar{y}$ is the average value of observation, and $\omega_{ij}$ is the spatial weight between observation i and j.

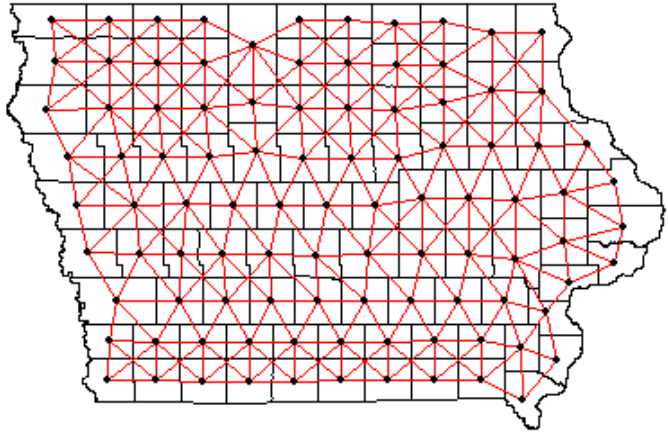

**Figure 2 Spatial neighbourhood structure for the counties in Iowa**

Negative Moran's *I* values indicate negative spatial autocorrelation, positive values indicate positive spatial autocorrelation, and zero indicates no spatial autocorrelation. The global Moran's I statistic of seatbelt usage is calculated for each month in 2022 using the "spdep" library package in Python with queen continuity spatial weights. In the queen's method, counties with shared border or vertex were considered as neighbors. When two counties are neighbors, the spatial weights are 1; otherwise, there are 0. The results are shown in Table 2.

Out of 12 months, 11 months have shown statistical significance at a 95% confidence level. Thus, statistically speaking, seatbelt percentages are not likely to be correlated at the county level in Iowa. However, looking at Figure 1, we see some spatial correlation in central Iowa around the red-shaded area, where the largest city of Iowa, Des Moines, is located. Another spatial grouping is also observed around southern Iowa. Although statistically, the spatial correlation came out to be insignificant, one of the reasons is the Normality assumption. Moran assumes the input data to be iid and normally distributed.

**Table 2: Global Moran's I statistics of Seatbelt Compliance in each month**

| Month | Moran's I | P-value |
|:-----:|:---------:|:-------:|
| 1 | 4.30327 | 8.41e-06* |
| 2 | 0.38701 | 3.49e-01 |
| 3 | 4.37216 | 6.15e-06* |
| 4 | 3.47753 | 2.53e-04* |
| 5 | 3.83048 | 6.39e-05* |
| 6 | 3.23529 | 6.08e-04* |
| 7 | 3.52585 | 2.11e-04* |
| 8 | 3.10867 | 9.40e-04* |
| 9 | 3.48215 | 2.49e-04* |
| 10 | 5.14411 | 1.34e-07* |
| 11 | 4.99327 | 2.97e-07* |
| 12 | 5.41587 | 3.05e-08* |

Figure 3 shows the histogram of the seatbelt usage percentage for the whole dataset. The skewness value for the data was -0.55, which tells us that the data is moderately skewed. Furthermore, studies like [36], [37] did a similar analysis at the county level in Pennsylvania, and the spatial correlation was insignificant. So, these trends can be site-specific, and without making any prior assumption, we considered the spatial correlation and included it in the modeling.

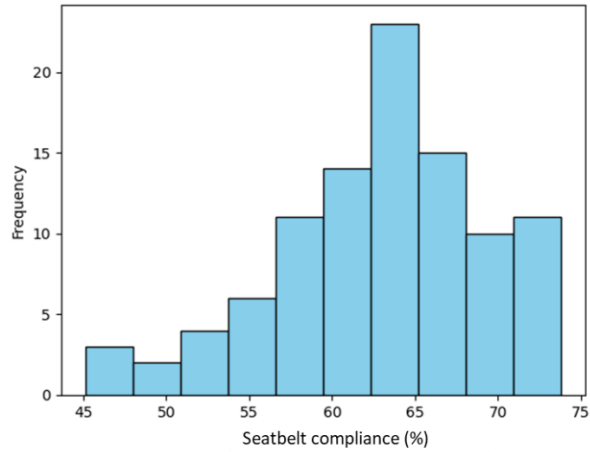

**Figure 3 Seatbelt Compliance distribution**

## METHODOLOGY

### Statistical framework

The framework uses a Bayesian hierarchical architecture, including spatial and temporal random effects. The model is explained in equations 2 and 3:

$$y_{it} \sim Beta(\lambda_{it})$$

$$logit(\lambda_{it}) = \alpha + \beta * X_{it} + \upsilon_i + v_i + \phi_t + \eta_{it}$$

Where $i$ is the county number from 1,2,…99; t is the month; $y_{it}$ is the seatbelt percentage compliance of the county $i$ in month $t$; $\lambda_{it}$ means the seatbelt percentage usage; $X_{it}$ is the covariate vector of the county $i$ in month t; $\upsilon_i$ is a structured spatial random effect of the county $i$; $v_i$ is the unstructured spatial random effect of the county $i$; $\phi_t$ is the temporal random effect in month t; and $\eta_{it}$ is the spatiotemporal interaction effect. Including spatial and temporal components helped us identify the underlying unobserved heterogeneity across counties and months.

### Integrated Nested Laplace Approximation (INLA)

Bayesian models are usually solved with Markov chain Monte Carlo (MCMC) simulations. However, MCMC becomes time-consuming when the models are complex. In this study, we employ both spatial and temporal effects, thereby making the model more complex. INLA proves to be much faster. For this analysis, we have used the R-INLA package in R to estimate the model parameters.

### Model Comparison and Checking

To assess the different models, we used measures such as the deviance information criterion (DIC) and the Akaike Information Criterion (AIC) to evaluate the adequacy of model fit.

DIC is used as a measure of assessing different Bayesian models and it is defined as

$$D = D(\bar{\theta}) + 2p_D = \bar{D} + p_D$$

where $D(\bar{\theta})$ is the deviance using the posterior mean values of the estimated parameters $(\bar{\theta})$, $D$ is the posterior mean of deviances, and $p_D$ is the effective number of parameters.

Root mean square error (RMSE) and mean absolute error (MAE) were also calculated to evaluate the adequacy of model fit.

In addition, RMSE and MEA we also calculated,

$$RMSE = \sqrt{\frac{1}{n_0} \Sigma_{j=1}^{n_0} (O_j - P_j)^2}$$

$$MAE = \frac{1}{n_0} \Sigma_{j=1}^{n_0} (O_j - P_j)^2$$

root mean square error (RMSE) and mean absolute error (MAE) were also calculated to evaluate the adequacy of model fit.

**Spatial Fraction Analysis**

To identify the contribution of the structured spatial effects $\sigma_v^2$ over the total marginal spatial variability $\sigma_v^2 + \sigma_v^2$ the spatial fraction is calculated, which is given by

$$frac_v = \frac{\sigma_v^2}{\sigma_v^2 + \sigma_v^2}$$

root mean square error (RMSE) and mean absolute error (MAE) were also calculated to evaluate the adequacy of model fit. When the spatial fraction is close to 1, the structured spatial effects account for most of the model's variability. Otherwise, the unstructured spatial random effects play the main role.

**Spatial effects and temporal effect assessment**

Multiple model configurations were tested to systematically evaluate the contribution of spatial and temporal components to model performance. Specifically, three individual models were developed: (1) a baseline model with no spatial or temporal effects, (2) a model incorporating only spatial effects, and (3) a model incorporating only temporal effects. These were then compared to the complete spatiotemporal model, which includes structured spatial and temporal dependencies. This comparative framework enables the isolation and assessment of the explanatory power of each component. All model specifications explored in this study are summarized in Table 3

**Table 3: Summary of models developed for the seatbelt usage analysis**

| No | Model Code | Spatial effect | Temporal effect | Spatio-temporal effect | Base model |
|----|------------|----------------|-----------------|------------------------|------------|
| 1 | $S_0T_0ST_0B$ | - | - | - | Beta |
| 2 | $S_{BYM}T_0ST_0B$ | BYM | - | - | Beta |
| 3 | $S_{BYM}T_LST_0B$ | BYM | Linear | | Beta |
| 4 | $S_{BYM}T_LST_1B$ | BYM | Linear | $\eta_{it}$ | Beta |
| 5 | $S_{BYM}T_{AR1}ST_1B$ | BYM | AR1 | $\eta_{it}$ | Beta |
| 6 | $S_{BYM}T_{RW1}ST_1B$ | BYM | RW1 | $\eta_{it}$ | Beta |

Note: 0, component not included; 1, component included; L, linear temporal; BYM, Besag-York-Mollie; AR1, 1st order autoregressive; RW1, 1st order random walk; "-" means non-existent.

## RESULTS AND DISCUSSIONS

To determine the most appropriate model for further analysis, four widely accepted evaluation metrics were employed: the Deviance Information Criterion (DIC), Conditional Predictive Ordinate (CPO), Root Mean Square Error (RMSE), and Mean Absolute Error (MAE). These metrics comprehensively understand model fit, predictive accuracy, and generalizability. Table 4 presents the values of these metrics for all six model configurations listed in Table 3. The most well-performing model can be identified and selected as the basis for subsequent interpretation and inference by comparing these indicators.

**Table 4 DIC, CPO, RMSE, and MAE values for all the models**

| No | Model Code | DIC | RMSE | MAE | CPO |
|----|------------|-----|------|-----|-----|
| 1 | $S_0T_0ST_0B$ | -3397.971 | 0.057956 | 0.043588 | 1.429898 |
| 2 | $S_{BYM}T_0ST_0B$ | -4181.596 | 0.038570 | 0.028013 | 1.758028 |
| 3 | $S_{BYM}T_LST_0B$ | -4200.391 | 0.037460 | 0.027185 | 1.762602 |
| 4 | $S_{BYM}T_LST_1B$ | -4201.454 | 0.037272 | 0.027051 | 1.762743 |
| 5 | $S_{BYM}T_{AR1}ST_1B$ | -4437.982 | 0.034167 | 0.023444 | 1.866161 |
| 6 | $S_{BYM}T_{RW1}ST_1B$ | -4437.770 | 0.034177 | 0.023442 | 1.866055 |

## Choice of Temporal Component

To evaluate the impact of temporal structure in crash modeling, three competing temporal components were tested within the Bayesian framework: a linear trend, a first-order autoregressive process (AR1), and a random walk of the first order (RW1). These were included in Models 3 through 6, all of which also incorporate structured spatial (BYM) effects and, in the case of Models 4-6, spatio-temporal interaction effects. Model 5, which includes an AR(1) temporal effect, yielded the lowest DIC (-4437.982) and the lowest RMSE (0.0342) and MAE (0.0234), indicating superior overall performance across both goodness-of-fit and predictive accuracy metrics. Although Model 6 (RW1) produced nearly identical results ($\Delta$DIC = +0.212), the AR(1) structure is marginally favored based on model selection criteria. Additionally, AR(1) provides a flexible yet parsimonious approach to capturing temporal autocorrelation, assuming that influences on crash risks in one year persist from the previous year with a gradually decaying memory, which is consistent with many real-world phenomena in transportation safety. Thus, the AR(1) process is selected as the preferred temporal structure based on numerical performance and modeling appropriateness.

## Including spatial, temporal, and spatio-temporal effects

The impact of including spatial, temporal, and spatio-temporal random effects is evident in the model comparisons. Moving from Model 1 (which lacks all structured effects) to Model 2 (which includes only spatial effects using the BYM formulation) results in a dramatic improvement in DIC (from -3397.97 to -4181.60), as well as significant reductions in RMSE and MAE. This confirms that spatial correlation across counties is a dominant source of variability in crash counts, likely reflecting regional clustering due to common infrastructure, demographic patterns, and policy environments. Further improvements were observed upon incorporating temporal trends (Models 3-6) and spatio-temporal interaction terms (Models 4-6). These models achieve lower

DIC values and better predictive calibration, as seen in the increasing Conditional Predictive Ordinate (CPO). Model 5, which incorporates structured spatial effects, an AR(1) temporal process, and independent and identically distributed (i.i.d.) spatio-temporal interactions, yields the best overall fit and predictive performance, suggesting that crash risks are influenced by both persistent regional factors and dynamic temporal shifts that vary across counties and years.

**Model Results**

Based on a comprehensive evaluation of all six Bayesia models, $S_{BYM}T_{AR1}ST_1B$ (Model 5) is selected as the optimal specification. This model incorporates a structured spatial effect (BYM) for county-level geographic clustering, a first-order autoregressive temporal process (AR1) to capture year-over-year dependence, and an independent and identically distributed (i.i.d.) spatio-temporal interaction term to model localized annual shocks.

Model 5 achieved the lowest DIC (-437.98), along with the best predictive accuracy (RMSE = 0.0342, MAE = 0.0234), and the highest mean CPO (1.866). These results underscore its ability to accurately represent the spatial and temporal complexity of fatal crash patterns in Iowa, while maintaining computational efficiency and interpretability. Table 5 presents the estimated parameters for the selected model,

**Table 5 Estimated parameters of the $S_{BYM}T_{AR1}ST_1B$ model**

| Parameter | Mean | SD | 0.025quant | 0.975quant |
|---|---|---|---|---|
| (Intercept) | -1.850 | 0.620 | -3.065 | -0.590 |
| Average Family Size | -0.120 | 0.180 | -0.470 | 0.215 |
| Total Household | 0.003 | 0.005 | -0.006 | 0.012 |
| Per capita income | 0.165 | 0.058 | 0.051 | 0.279 |
| Unemployment rate | 0.004 | 0.016 | -0.032 | 0.041 |
| BPL Families | 0.012 | 0.009 | -0.006 | 0.030 |
| Proportion of Young | 3.980 | 0.940 | 2.130 | 5.800 |
| Proportion of Female | 4.550 | 0.980 | 2.650 | 6.510 |
| Proportion of Male | -1.050 | 2.200 | -5.310 | 3.210 |
| VMT | 0.021 | 0.006 | 0.009 | 0.033 |

After removing the statistically insignificant variables from the model, only two covariates, per capita income and vehicle miles traveled (VMT), were retained. The final model results for $S_{BYM}T_{AR1}ST_1B$ both are statistically significant predictors of seatbelt compliance, with 95% credible intervals that exclude zero, confirming their positive association with the outcome. The estimated result is presented in Table 6

**Table 6 Estimated parameters of the $S_{BYM}T_{AR1}ST_1B$ model with VMT and Per Capita Income**

| Parameter | Mean | SD | 0.025quant | 0.975quant |
|---|---|---|---|---|
| (Intercept) | -0.210 | 0.090 | -0.387 | -0.032 |
| Per Capita Income | 0.165 | 0.058 | 0.051 | 0.279 |
| VMT | 0.021 | 0.006 | 0.009 | 0.033 |

## POST-MODEL SPATIAL DIAGNOSTIC: RESIDUAL SPATIAL AUTOCORRELATION

To comprehensively assess the adequacy and robustness of the final model, a set of post-model evaluation techniques was conducted. These diagnostics aim to verify that the model has effectively accounted for spatial and temporal structure in the data, and to ensure that residual patterns do not exhibit systematic biases. Three key evaluations were performed: Moran's I test on residuals to detect any remaining spatial autocorrelation, examining the PIT plot to assess predictive calibration, and visualizing the structured spatial effects to explore geographic patterns in seatbelt compliance. The subsequent subsections provide a detailed discussion of them.

### Residual Spatial Autocorrelation

To assess whether the final model fully captured the spatial structure, a Moran's I test was applied to the model residuals. The results in Table 7 show that, in most cases, Moran's I values were close to zero with non-significant p-values, indicating little to no spatial autocorrelation. However, statistically significant clustering of residuals was observed in a few instances (e.g., Moran's I = 1.88, p = 0.0297 and 2.86, p = 0.0021), suggesting that minor residual spatial dependence persists in localized patterns.

**Table 7 Moran's I test result for the residual of the $S_{BYM}T_{AR1}ST_1B$ model**

| Month | Moran's I | p-value |
|---|---|---|
| 1 | 1.884 | 0.0297 |
| 2 | -0.867 | 0.8071 |
| 3 | 1.852 | 0.0320 |
| 4 | -0.337 | 0.6318 |
| 5 | -0.012 | 0.5049 |
| 6 | 0.155 | 0.4384 |
| 7 | 0.648 | 0.2585 |
| 8 | 1.296 | 0.0975 |
| 9 | -0.067 | 0.5268 |
| 10 | 2.856 | 0.0021 |

The corresponding residual map in Figure 4 shows that the residual values across counties are generally small and symmetrically distributed around zero, with most counties falling in the neutral range (-0.1 to 0.1). Only a few counties show higher residuals (±0.2 to ±0.3), and no clear geographic clustering is visible. These results suggest that the model has captured the dominant spatial structure, and any remaining spatial correlation is minimal and localized.

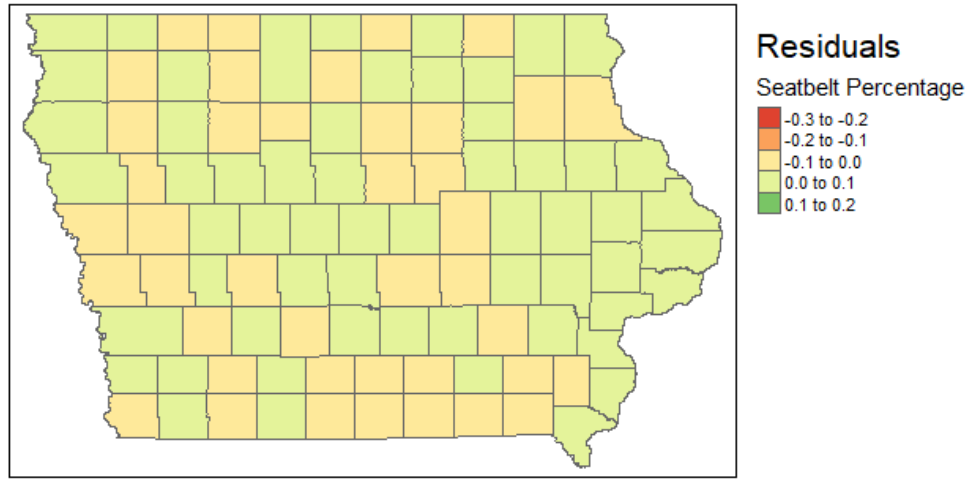

**Figure 4 County-level spatial distribution of residuals from the $S_{BYM}T_{AR1}ST_1B$   Model.**

**Evaluation of Predictive Calibration Using PIT Plot**

As part of the post-model diagnostic process, a Probability Integral Transform (PIT) plot was used to evaluate the predictive calibration of the final spatio-temporal model, which incorporates a structured spatial effect (BYM) and an AR(1) temporal component. A PIT histogram provides insight into how well the model's predicted distributions match the observed data. Ideally, a uniform distribution of PIT values indicates good calibration. Figure 5 shows that the adjusted PIT values exhibit a reasonable uniformity, with only minor deviations across certain intervals. This pattern suggests that the model's predictive distribution aligns well with the observed data across counties and periods. Small fluctuations are common in complex hierarchical models and do not necessarily indicate serious misspecification. Overall, the PIT plot confirms that the final model, with its AR(1) temporal structure, demonstrates adequate predictive performance and is well-calibrated for analyzing seatbelt compliance patterns in Iowa.

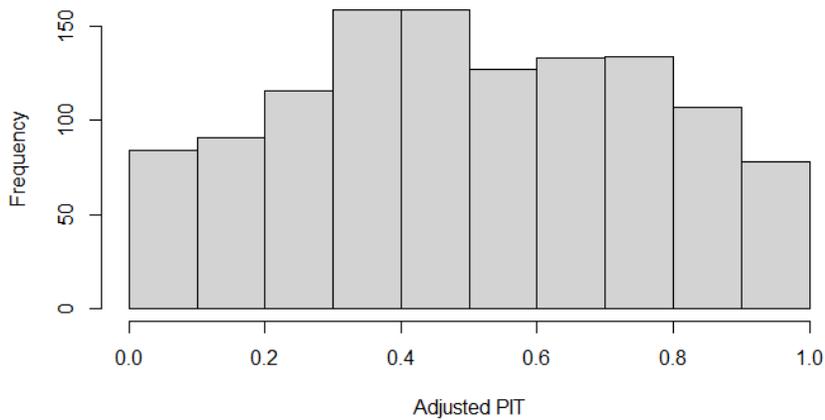

**Figure 5 Histogram of the adjusted PIT values of the $S_{BYM}T_{AR1}ST_1B$   model**

**Structured Spatial Effects and County-Level Seatbelt Compliance**

The posterior means of the structured spatial effects were exponentiated and mapped across Iowa counties to explore the spatial distribution of seatbelt compliance after accounting for covariates and random effects (Figure 6). After adjusting for socio-economic and temporal factors, these values, interpreted as relative compliance, represent how each county compares to the statewide average.

In Figure 6, counties where $e(v_i) > 1$ (i.e., higher seatbelt compliance) are shaded in green, while counties with $e(v_i) < 1$ (i.e., lower seatbelt compliance) are shown in red. This visual highlights areas where compliance lags behind the statewide baseline. Several red-shaded counties are observed across northern and southern Iowa, indicating lower seatbelt usage than expected, even after accounting for socio-economic and temporal factors. In contrast, green-shaded counties, particularly in the north and south regions, reflect relatively higher compliance, suggesting more effective seatbelt usage behavior or possibly more vigorous enforcement and outreach.

This spatial heterogeneity suggests that geographic differences persist even after adjusting for known covariates, potentially driven by local policies, enforcement intensity, cultural norms, or unobserved behavioral factors. These findings highlight the importance of targeted, location-specific interventions, particularly in counties that consistently demonstrate lower compliance rates. Identifying such spatial trends is crucial for prioritizing outreach and enforcement programs to improve statewide seatbelt use.

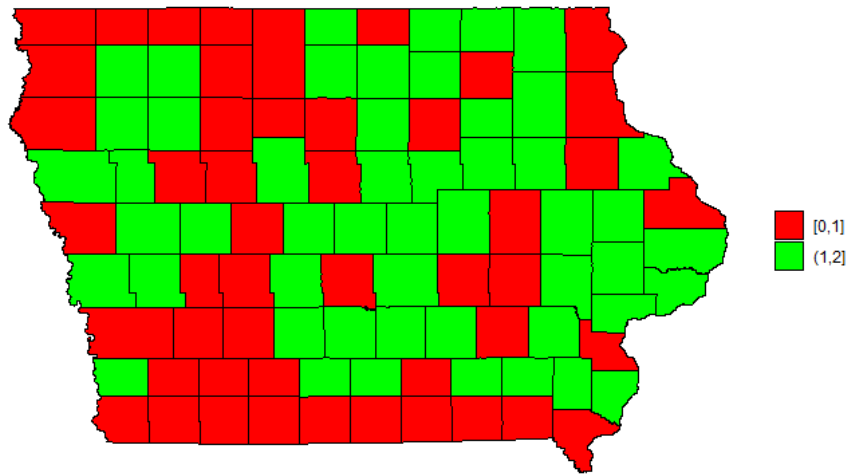

**County-Level Seatbelt Compliance**
Counties with exp($u_i$) ≥ 1 have higher compliance; exp($u_i$) < 1 have lower compliance

[0,1]
(1,2)

**Figure 6: Exponential posterior means of the structured spatial effect**

**CONCLUSION AND FUTURE RESEARCH**

This study presents a novel application of spatiotemporal analysis for seat belt compliance by leveraging telematics data, an approach that is relatively unexplored in the existing literature. By utilizing telematics data, this research demonstrates a cost-effective and scalable approach to obtaining highly granular compliance measurements across all 99 counties in Iowa for 2022. The rich spatial and temporal resolution of telematics-derived "driver event" data enabled the development and comparison of various beta-regression models incorporating spatial, temporal, and combined spatio-temporal random effects. Among the explanatory variables examined, only vehicle miles traveled (VMT) and per capita income were statistically significant predictors of seat belt compliance. The findings highlight the critical role of both spatial and temporal dependencies in accurately modeling compliance patterns, underscoring the importance of incorporating such complex structures for improved policy assessment and targeted intervention design.

For future work, extending the analysis over multiple years or seasonal periods could reveal long-term trends and changes in seat belt use, providing deeper insights for evaluating the sustained impact of safety campaigns and enforcement policies. Additionally, telematics data offers unparalleled granularity but is limited by variable market penetration rates, potential demographic biases in the instrumented fleet, and data privacy constraints. Further research should focus on validating telematics-based compliance estimates against diverse ground-truth sources, addressing sampling representativeness, and integrating emerging data streams, such as smartphone telematics or infrastructure sensor networks. Exploring advanced spatio-temporal modeling techniques that better capture nonstationary or non-linear effects may also enhance our understanding of the complex drivers influencing seat belt compliance over space and time.

## ACKNOWLEDGEMENT

The authors would like to acknowledge the Iowa DOT for providing access to the $3rd$ party telematics data used in this analysis. This work builds off research conducted by the Institute for Transportation at Iowa State University in Ames, IA, in support of the Iowa DOT's Systems Operations Division. The contents of this paper reflect the views of the authors and do not necessarily reflect the official views or policies of the sponsoring organization. AI tools, such as Grammarly, were employed solely for language editing and manuscript refinement to enhance clarity and readability, without influencing the scientific content, data analysis, or interpretation.

## AUTHORS CONTRIBUTION

The authors confirm their contribution to the paper as follows: A. Dumka conceived the study, carried out the analysis, drafted, and finalized the paper. R. Kandiboina helped refine the draft. S. Knickerbocker, J. Wood, and A. Sharma were co-advisors who fine-tuned the study. N. Hawkins was also a committee member who contributed to fine-tuning the study. All authors reviewed the results and approved the final version of the manuscript.